\documentclass[
	conference,
	final,
	10pt,
	letterpaper,
	oneside,
	twocolumn
]{IEEEtran}
\IEEEoverridecommandlockouts

\usepackage[sort,compress]{cite}

\usepackage{amsmath,amssymb,amsfonts}

\usepackage{algorithm, algorithmic}

\usepackage{booktabs}

\usepackage{url}

\usepackage{hyperref}
\hypersetup{
hidelinks,
pdfauthor={Ryan Tatton},
pdfcreator={Ryan Tatton},
pdfproducer={Ryan Tatton},
pdftitle={ShareTrace: Contact Tracing with the Actor Model},
pdfkeywords={actor model} {Contact tracing} {COVID-19} {Graph algorithms} {Graphical models} {Message passing} {Temporal network},
pdffitwindow=false,
pdfstartview={FitH},
}

\usepackage{cleveref}
\crefname{figure}{fig.}{figs.}
\crefname{enumi}{step}{steps}
\crefformat{equation}{(#2#1#3)}

\usepackage{hyphenat}

\usepackage[caption=false,font=footnotesize]{subfig}

\usepackage{tikz}
\usetikzlibrary{bayesnet, plotmarks, calc}

\usepackage{pgfplots}
\usepgfplotslibrary{groupplots, statistics}
\pgfplotsset{
    compat=1.15,
    only if/.style args={entry of #1 is #2}{
        /pgfplots/boxplot/data filter/.code={
            \edef\tempa{\thisrow{#1}}
            \edef\tempb{#2}
            \ifx\tempa\tempb
            \else
                
            \fi
        }
    }
}
\newcommand{\tbuf}{\beta}
\newcommand{\trate}{\alpha}
\newcommand{\scoeff}{\gamma}
\newcommand{\lookback}{L}
\newcommand{\nactors}{K}
\newcommand{\timeout}{T}
\newcommand{\vsym}{r}
\newcommand{\tsym}{t}
\newcommand{\msym}{m}
\newcommand{\graph}{G}
\newcommand{\iactor}{k}
\DeclareMathOperator{\argmax}{\arg\max}
\DeclareMathOperator{\nhood}{ne}
\DeclareMathOperator{\mrr}{mrr}
\newcommand{\init}{\msym_0}
\newcommand{\curr}{\vsym}
\newcommand{\ival}{\vsym_0}
\newcommand{\itime}{\tsym_0}
\newcommand{\msg}[2]{\msym_{#1 \rightarrow #2}}
\newcommand{\mval}[2]{\vsym_{#1 \rightarrow #2}}
\newcommand{\mtime}[2]{\tsym_{#1 \rightarrow #2}}
\newcommand{\reach}{\msym}
\newcommand{\estreach}{\hat{\msym}}
\newcommand{\card}[1]{\left\vert #1 \right\vert}
\newcommand{\vpath}[2]{#1 \rightarrow #2}
\newcommand{\contacts}{\mathcal{C}}
\newcommand{\rscores}{\mathcal{R}}
\newcommand{\scores}{\mathcal{S}}
\newcommand{\edges}{\mathcal{E}}
\newcommand{\variables}{\mathcal{V}}
\newcommand{\factors}{\mathcal{F}}
\newcommand{\actors}{\mathcal{G}}
\newcommand{\pathsym}{\mathcal{P}}

\begin{document}
\bstctlcite{BSTcontrol}

\title{ShareTrace: Contact Tracing with the Actor Model\thanks{Research reported in this paper was partly supported by the National Science Foundation (NSF) under grant number NSF CCF 2200255 and Cisco Research University Funding grant number 2800379.}}

\author{%
\IEEEauthorblockN{Ryan Tatton\IEEEauthorrefmark{1}, Erman Ayday\IEEEauthorrefmark{2}}
\IEEEauthorblockA{%
Department of Computer and Data Sciences\\
Case Western Reserve University\\
Cleveland, Ohio, USA\\
\IEEEauthorrefmark{1}ryan.tatton@case.edu, \IEEEauthorrefmark{2}erman.ayday@case.edu}%
\and
\IEEEauthorblockN{Youngjin Yoo}%
\IEEEauthorblockA{%
Department of Design and Innovation\\
Case Western Reserve University\\
Cleveland, Ohio, USA\\
youngjin.yoo@case.edu}%
\and
\IEEEauthorblockN{%
Anisa Halimi\IEEEauthorrefmark{3}%
\thanks{\IEEEauthorrefmark{3}Work was completed while attending Case Western Reserve University.}}%
\IEEEauthorblockA{%
IBM Research Europe\\
Dublin, Ireland\\
anisa.halimi@ibm.com}}%

\maketitle

\begin{abstract}
Proximity\hyp{}based contact tracing relies on mobile\hyp{}device interaction to estimate the spread of disease. ShareTrace is one such approach that improves the efficacy of tracking disease spread by considering direct and indirect forms of contact. In this work, we utilize the actor model to provide an efficient and scalable formulation of ShareTrace with asynchronous, concurrent message passing on a temporal contact network. We also introduce message reachability, an extension of temporal reachability that accounts for network topology and message\hyp{}passing semantics. Our evaluation on both synthetic and real\hyp{}world contact networks indicates that correct parameter values optimize for algorithmic accuracy and efficiency. In addition, we demonstrate that message reachability can accurately estimate the risk a user poses to their contacts.
\end{abstract}


\section{Introduction}
ShareTrace is a privacy\hyp{}preserving contact\hyp{}tracing solution \cite{Ayday2021}. Unlike other approaches that rely on device proximity to detect human interaction, ShareTrace executes iterative message passing on a factor graph to estimate a user's marginal posterior infection probability (MPIP). To indicate its similarity to belief propagation, we refer to the ShareTrace algorithm as \emph{risk propagation}. By considering direct \emph{and} indirect contacts, \cite{Ayday2021} demonstrates that risk propagation is more effective than other proximity\hyp{}based methods that only consider the former.

Building upon the efforts by \cite{Ayday2021}, we provide an efficient and scalable formulation of risk propagation\footnote{\url{https://github.com/share-trace}} that utilizes asynchronous, concurrent message passing on a temporal contact network \cite{Holme2012, Holme2015}. Our application of message passing on a temporal network differs from previous epidemiological works. Notably, we use a temporal network to infer a user's infection \emph{risk}, unlike its typical usage for modeling the spreading dynamics of the infection itself \cite{Danon2011, Karrer2010, Lokhov2014, Pastor-Satorras2015, Koher2019, Li2021, Zino2021}. We also introduce message reachability, an extension of temporal reachability that accounts for network topology and message\hyp{}passing semantics. Our formulation of risk propagation aligns with its distributed extension \cite{Ayday2021}, which has connections to the actor model \cite{Baker1977, Agha1986} and the ``think\hyp{}like\hyp{}a\hyp{}vertex'' model of graph\hyp{}processing algorithms \cite{McCune2015}. Our evaluation aims to quantify the efficiency and scalability of this new formulation of risk propagation, as well as validate the accuracy of message reachability. To keep the scope of this work focused, we defer to \cite{Ayday2021} on the privacy and security aspects of ShareTrace.

\section{Related Work}
Since the beginning of the COVID\hyp{}19 pandemic, there has been a copious amount of research in mobile contact tracing solutions, most notably being the joint effort by Apple and Google \cite{AppleGoogle}. External reviews and surveys provide extensive comparison of existing solutions through the lenses of privacy, security, ethics, adversarial models, data management, scalability, interoperability, and more. References \cite{Ahmed2020, Martin2020} provide thorough reviews of existing contact tracing solutions with discussion of the techniques, privacy, security, and adversarial models. The former offers additional detail on the system architecture (i.e., centralized, decentralized, and hybrid), data management, and user concerns of existing solutions. Other notable reviews with similar discussion include \cite{Wen2020, Raskar2020, Cho2020, Dar2020, Lucivero2020}. Reference \cite{Kuhn2021} provides a formal framework for defining aspects of privacy for proximity\hyp{}based contact tracing.

\section{Proposed Scheme}\label{sec:risk-prop}
\subsection{Preliminaries}
We assume a system model in which each user owns a mobile device that has device\hyp{}proximity detection (e.g., Bluetooth); and that proximal interactions between devices subsequently allow them, or a digital proxy thereof, to exchange messages over several days.

In risk propagation, computing infection risk is an inference problem in which the task is to estimate a user's MPIP. We derive prior infection probability from user symptoms \cite{Menni2020}, so we refer to it, along with the time of its computation, as a \emph{symptom score}. Because the posterior infection probability accounts for contact with other users, we call it an \emph{exposure score}. In general, a \emph{risk score} $(\vsym, \tsym)$ is a timestamped infection probability where $\vsym \in [0, 1]$ is the \emph{value} of the risk score and $\tsym \geq 0$ is the \emph{time} of its computation.

Computing the full joint probability distribution is intractable as it scales exponentially with the number of users. To circumvent this challenge, risk propagation uses message passing on a factor graph to efficiently compute the MPIP. Formally, let $\graph = (\variables, \factors, \edges)$ be a factor graph where $\variables$ is the set of variable vertices, $\factors$ is the set of factor vertices, and $\edges$ is the set of edges incident between them \cite{Kschischang2001}. A \emph{variable vertex} $v \in \variables$ is a random variable that represents the probability of infection for a user. For this reason, we use ``user'' and ``variable vertex'' interchangably in this work. A \emph{factor vertex} $f(u, v) \in \factors$ represents contact between users $u, v \in \variables$ such that $f(u, v)$ is adjacent to them. While belief propagation aims to maximize the full joint distribution \cite{Bishop2006}, risk propagation aims to maximize individual MPIPs \cite{Ayday2021}.

A \emph{message} $\msg{u}{v} = \{(\vsym, \tsym),\ldots\}$ sent from vertex $u$ to vertex $v$ is a nonempty set of risk scores. We assume that contact has a nondecreasing effect on a user's infection probability. Thus, risk propagation is similar to the max\hyp{}sum algorithm in that each variable vertex maintains the value of the maximum risk score it receives \cite{Bishop2006}.

The only purpose of a factor vertex is to compute and relay messages between variable vertices. Thus, we can apply one\hyp{}mode projection such that variable vertices $u, v$ are adjacent if the factor vertex $f(u, v)$ exists \cite{Zhou2007}. To send a message to variable vertex $v$, variable vertex $u$ applies the computation that was associated with the factor vertex $f(u, v)$. This modification differs from the distributed extension of risk propagation \cite{Ayday2021} in that we do not duplicate factor vertices and messages. By storing the contact time between users on the edge incident to their variable vertices, this modified topology is the \emph{contact-sequence} representation of a \emph{contact network}, a kind of temporal network in which vertices represent people and edges indicate that two people came in contact:
\begin{displaymath}
	\contacts = \{(u, v, t) \mid u, v \in \variables; u \neq v; \tsym \geq 0\},
\end{displaymath}
where a triple $(u, v, t)$ is called a \emph{contact} \cite{Holme2012}. Specific to risk propagation, $\tsym$ is the time at which users $u$ and $v$ \emph{most recently} came in contact.

We utilize the actor model to achieve scalable performance \cite{Baker1977, Agha1986}. Let $\nactors$ be the number of actors, where each actor is a subnetwork $\graph_{\iactor} \in \actors$ of users that is induced by a partitioning algorithm \cite{Buluc2016}. Formally, we apply a surjective function $\sigma: \variables \rightarrow \actors$ that maps each user to a subnetwork actor. Actors communicate via message passing. Typically, due to the underlying implementation, inter\hyp{}actor communication is slower than intra\hyp{}actor computation, so using an algorithm that minimizes communication complexity between actors is key to maximizing performance.

We associate with each actor an identifier, its \emph{mailing address}, and a buffer, its \emph{remote mailbox}, for storing received messages. In practice, each actor also has a \emph{local mailbox} that it is uses to manage communication between its own users. This local mailbox incurs less overhead than the remote mailbox since the latter typically involves the usage of concurrent primitives. To send a message, an actor must know the mailing address of the receiving actor and the identity of the receiving user. If the mailing address of the sending actor is the same as the receiving actor, then the message is placed in its local mailbox. Otherwise, the message is placed in the remote mailbox of the receiving actor. In addition to maintaining the state of its subnetwork, an actor also keeps a mapping between mailing addresses and remote mailboxes for all other actors, because it is unknown with which actors it needs to communicate before partitioning the network.

\subsection{Algorithms}\label{sec:algorithms}
\Cref{alg:rp-main} defines the main message-passing procedure. We constrain the set of initial risk scores $\scores$ (resp. contacts $\contacts$) to those that were computed (resp. occurred) within the last $\lookback = 14$ days, which assumes that a risk score (resp. contact) has finite relevance. Note that the initial risk scores of a user $u$, denoted $\scores(u)$, includes the exposure scores from the last $\lookback$ days and its most recently computed symptom score.
\begin{algorithm}[tbh]
\begin{enumerate}
    \item Create the network $\graph$: for each contact $(u, v, \tsym) \in \contacts$, add an edge between users $u, v$ and store the contact time $\tsym$.
    \item Partition $\graph$ into $\nactors$ disjoint actor subnetworks w.r.t. a partitioning function $\sigma$.
    \item Partition the initial risk scores $\scores$ w.r.t. $\sigma(\graph)$.
    \item Send $\scores_{\iactor}$ to $\graph_{\iactor}$ for each $\iactor = 1, 2, \ldots, \nactors$. \label{item:send}
    \item Collect all exposure scores: $\rscores = \cup_{\iactor} \rscores_{\iactor}$. \label{item:collect}
\end{enumerate}
\caption{Risk Propagation, Main.}
\label{alg:rp-main}
\end{algorithm}

\Cref{alg:rp-actor} describes the behavior of an actor. As in \cite{Ayday2021}, we assume that risk transmission is incomplete by applying a transmission rate of $\trate = 0.8 \in (0, 1)$ \cite{Hamner2020}. Step \ref{item:on-next} follows from belief propagation in that we marginalize over the factor $f(u, v)$ \cite{Ayday2021}. Because message passing is concurrent and asynchronous, we cannot rely on a global iteration or an inter\hyp{}iteration difference threshold as stopping criteria, as in \cite{Ayday2021}. While convenient, such criteria require synchronization which can degrade performance \cite{Han2015}.
\begin{algorithm}[tbh]
\begin{enumerate}
    \item Upon receiving $\scores_{\iactor}$, for each user $u \in \graph_{\iactor}$, let \label{item:attrs}
        \begin{enumerate}
            \item $\init(u) = (\ival(u), \itime(u))$ be its initial message, i.e., its maximum risk score, scaled by $\trate$; and
            \item $\curr(u)$ be its current value; initially, $\max\left(\scores(u)\right)$.
        \end{enumerate}
    \item For each user $u \in \graph_{\iactor}$, compute and send the message $\msg{u}{v}$ using $\scores(u)$, for each neighbor $v \in \nhood(u)$. \label{item:init-msg}
    \item While a message has been received within $\timeout$ seconds, \label{item:while}
        \begin{enumerate}
            \item Receive $\msg{v}{u}$ s.t. $u \in \graph_{\iactor}$ and $v \in \nhood(u)$.
            \item Update user $u$: $\curr(u) \leftarrow \max(\mval{v}{u}, \curr(u))$ \label{item:on-initial}.
            \item For each $v' \in \nhood(u) \setminus v$, compute and send $\msg{u}{v'}$.\label{item:on-next}
        \end{enumerate}
    \item Collect exposure scores: $\rscores_{\iactor} = \{(\curr(u), t_{\text{now}}) \mid u \in \graph_{\iactor}\}$.
\end{enumerate}
\caption{Risk Propagation, Actor (Main).}
\label{alg:rp-actor}
\end{algorithm}

\Cref{alg:rp-msg} describes how to compute and send a message. As indicated by \cref{item:init-msg} of \Cref{alg:rp-actor}, the message $\msg{v}{u}$ in \cref{item:filter} is initially the risk scores of user $u$. Thus, we do not apply \cref{item:filter} for these ``self-messages.'' For all subsequent messages, $u \neq v$ and $\msg{v}{u}$ is a singleton that contains the risk score sent from neighbor $v$. For a singleton message $\msg{v}{u}$, we refer to the value (resp. time) of the contained risk score as $\mval{v}{u}$ (resp. $\mtime{v}{u}$). Furthermore, we use ``risk score'' and ``message'' interchangably.

In \cref{item:filter}, we include a \emph{time buffer} of $\tbuf = 2 \geq 0$ days to account for the disease incubation period or delayed symptom reporting. We assume that all risk scores with a time later than the buffered contact time are irrelevant. Assuming that we run risk propagation at least every $\tbuf$ days, it is unnecessary to persist contacts older than $\tbuf$ days. For a given user $u$ and neighbor $v$, it is impossible for $u$ to send $v$ a risk score higher in value than what it previously sent if it has been more than $\tbuf$ days after their most recent contact time. In other words, the MPIP of user $v$ will already account for any risk score of user $u$ after $\tbuf$ days of coming in contact. In this way, we can further improve the efficiency of risk propagation by reducing the communication overhead.

The final aspect of \Cref{alg:rp-msg} is to determine if we should send the computed message. Because we only use contact time as a filter to determine which risk scores to consider, we only need to compare the most recent contact time in \cref{item:filter}. That is, given contact times $\tsym_1 \leq \tsym_2$ and risk score time $\tsym \leq \tsym_1 + \tbuf$, it follows that $\tsym \leq \tsym_2 + \tbuf$. This avoids storing and comparing multiple contact times, as suggested by \cite{Ayday2021}.

The intent of sending a message is to update the value of other users in the network. Previous work \cite{Ayday2021} would send a ``null'' message with a value of 0 if $\msg{v}{u}' = \emptyset$ in \cref{item:filter}. However, sending a risk score to a user that has a lower value is neither useful nor efficient. It also holds that for a sufficiently old risk score, propagating it may affect the value of an indirect contact, even if the risk score is relatively low. We combine both of these aspects into a heuristic that allows us to parametrize the trade\hyp{}off between accuracy and efficiency in an asynchronous, concurrent setting. Let $\scoeff \in [0, 1]$ be the \emph{send coefficient} such that we only send a message $\msg{u}{v}$ if $\mval{u}{v} \geq \scoeff \cdot \ival(u)$. In addition to comparing the value, we must also compare its time to the initial message. Assuming a message satisfies the value condition, then a newer message is less likely to be propagated. Hence, it is only useful to send a message if it is at least as old as the initial message. This send condition is expressed in \cref{item:send-condition}. Because we scale the value of a risk score by the transmission rate, it exponentially decreases as it propagates away from the source user with a rate constant of $\log(\trate)$. Assuming a finite number of users, a positive send coefficient guarantees that we will propagate a risk score finitely many times. Therefore, for $\scoeff > 0$, the send condition will eventually cause actors to stop passing messages, thus terminating risk propagation.
\begin{algorithm}[tbh]
\begin{enumerate}
   	\item Consider only the risk scores in the message $\msg{v}{u}$ that may have been transmitted: \label{item:filter}
	    \begin{displaymath}
            \msg{v}{u}' \leftarrow \left\{(\vsym, \tsym) \mid \tsym \leq \tsym_{uv} + \tbuf \right\}.
	    \end{displaymath}
   	\item Compute the time difference for each remaining score: \label{item:delta}
	    \begin{displaymath}
   	        \Delta \leftarrow \{(\vsym, \tsym, \delta) \mid \delta = \min(\tsym - \tsym_{uv}, 0)\}.
	    \end{displaymath}
   	\item Compute the maximum weighted message: \label{item:argmax}
	    \begin{displaymath}
         \msg{u}{v'} \leftarrow \underset{\msym \in \Delta}{\argmax} \left\{\log(\vsym) + \delta / \tau \right\}.
	    \end{displaymath}
   	\item Scale by the transmission rate: $\mval{u}{v'} \leftarrow \trate \cdot \mval{u}{v'}$.
   	\item Send $\msg{u}{v'}$ if $\mval{u}{v'} \geq \scoeff \cdot \ival(u)$ and $\mtime{u}{v'} \leq \itime(u)$. \label{item:send-condition}
\end{enumerate}
\caption{Risk Propagation, Actor (Message).}
\label{alg:rp-msg}
\end{algorithm}

\subsection{Message Reachability}\label{sec:reachability}
A fundamental concept in reachability analysis on a temporal network is a \emph{time\hyp{}respecting path}: a contiguous sequence of contacts with nondecreasing time. Thus, vertex $v$ is \emph{temporally reachable} from vertex $u$ if there exists a time\hyp{}respecting path from $u$ to $v$ \cite{Holme2012}. Generally, a message\hyp{}passing algorithm defines a set of constraints that determine when a vertex sends a message. Even if operating on a temporal network, those constraints may not require temporal reachability. As a dynamic process, message passing on a time\hyp{}varying network necessitates a more general definition of reachability that can account for network topology \emph{and} message\hyp{}passing semantics \cite{Barrat2013}. Formally, the \emph{message reachability from vertex $u$ to vertex $v$} is the number of edges along the \emph{shortest path} $\pathsym = \vpath{u}{v}$ that satisfy the message\hyp{}passing constraints,
\begin{displaymath}
	\reach(u, v) = \sum_{(i, j) \in \pathsym} f(u, i, j, v),
\end{displaymath}
where $f(u, i, j, v) = 1$ if all constraints are satisfied and $f(u, i, j, v) = 0$ otherwise.
Vertex $v$ is \emph{message reachable} from vertex $u$ if there exists a shortest path $\pathsym = \vpath{u}{v}$ such that $\reach(u, v) = \card{\pathsym}$; such a path is \emph{message respecting}. The \emph{message reachability of vertex $u$} is the maximum message reachability from vertex $u$: $\reach(u) = \max \{\reach(u, v) \mid v \in \variables \}$. Temporal reachability concepts, such as the influence set, source set, and reachability ratio \cite{Holme2012}, can be extended to message\hyp{}passing contexts by defining them in terms of a message\hyp{}respecting path, rather than a time\hyp{}respecting path.

For risk propagation, the message reachability of a user is the length of the longest shortest path over which their initial risk score can be passed. Using the \emph{Heaviside step function} $H(x) = \mathbf{1}_{x \geq 0}$, message reachability is defined as
\begin{equation}\label{eq:reach}
    \reach(u) = \underset{\pathsym}{\max} \left\{\sum_{(i, j) \in \pathsym} f_c(u, i, j) \cdot f_{\vsym}(u, i) \cdot f_{\tsym}(u, i) \right\},
\end{equation}
where users are enumerated $0, 1, \ldots, \card{\pathsym} - 1$; and
\begin{align}
  	f_c(u, i, j) &= H(\tsym_{ij} + \tbuf - \itime(u)) \label{eq:contact-const} \\
    f_r(u, i) &= H(\trate^i \cdot \ival(u) - \scoeff \cdot \ival(i)) \label{eq:val-const}\\
    f_t(u, i) &= H(\itime(i) - \itime(u)) \label{eq:time-const}
\end{align}
are the contact\hyp{}time, value, and time constraints, respectively. Because we constrain risk scores to be at most $\lookback \geq \tbuf$ days old, $\reach(u) \geq 1$ for any non\hyp{}isolated user $u$. We can find the value of \Cref{eq:reach} by applying an augmented shortest\hyp{}path algorithm \cite{Johnson1977} such that we start at user $u$ and iteratively propagate its initial message $\init(u)$. By relaxing \Cref{eq:contact-const} and \Cref{eq:time-const}, we can define an upper bound on \Cref{eq:reach} with \Cref{eq:val-const}. For some reachable user $v$, the \emph{estimated message reachability of vertex $u$ to vertex $v$} is
\begin{equation}\label{eq:estreach}
	\estreach(u, v) = \log_{\trate}\left\{\scoeff \cdot \frac{\ival(v)}{\ival(u)} \right\}.
\end{equation}

\nameCref{eq:reach} \Cref{eq:reach} helps quantify the communication complexity of a given message\hyp{}passing algorithm on a temporal network. Specific to risk propagation, message reachability estimates the size of the induced subnetwork (i.e., set of users) that is impacted by a user's infection risk. \nameCref{eq:reach} \Cref{eq:estreach} indicates that a lower send coefficient will typically result in higher message reachability, at the cost of computing and passing redundant messages (i.e., messages that do not change the exposure score of another user). It also allows us to quantify the effect of the transmission rate. However, unlike the send coefficient that should be optimized, the transmission rate should be derived from epidemiology to quantify infectivity.

\section{Evaluation}\label{sec:evaluation}
\subsection{Experimental Design}
Risk propagation requires a partitioning algorithm, as described in \Cref{alg:rp-main}. We configured the METIS algorithm \cite{Karypis1998} to use $k$\hyp{}way partitioning with a load imbalance factor of 0.2 and to attempt contiguous partitions that have minimal inter\hyp{}partition connectivity. We applied 10 iterations of refinement during each stage of the uncoarsening process and used the best of 3 cuts.

\subsubsection{Synthetic Networks}\label{sec:synthetic-eval}
We evaluated the scalability and efficiency of risk propagation on random geometric graphs (RGGs) \cite{Dall2002}, benchmark graphs (LFRGs) \cite{Lancichinetti2008}, and clustered scale\hyp{}free graphs (CSFGs) \cite{Holme2002}. Together, these graphs demonstrate some aspects of community structure \cite{Fortunato2010} which ensured a fair performance measurement. When constructing a RGG, we set the radius to $r(n) = \min \left(1, 0.25^{\log_{10}(n) - 1}\right)$, where $n$ is the number of users. This allowed us to scale the size of the network while maintaining reasonable density. To generate LFRGs, we used the following parameter values: mixing parameter $\mu = 0.1$, degree power\hyp{}law exponent $\gamma = 3$, community\hyp{}size power\hyp{}law exponent $\beta = 2$, degree bounds $(k_{\min}, k_{\max}) = (3, 50)$, and community\hyp{}size bounds $(s_{\min}, s_{\max}) = (10, 100)$. These align with the suggestions by \cite{Lancichinetti2008} in that $\gamma \in [2, 3]$, $\beta \in [1, 2]$, $k_{\min} < s_{\min}$, and $k_{\max} < s_{\max}$. To build CSFGs, we added $m = 2$ edges for each new user and used a triad\hyp{}formulation probability of $P_t = 0.95$. For all networks, we removed self\hyp{}loops and isolated vertices.

The following describes our data generation process. Let $p$ be the probability of a user being ``high risk'' (i.e., $\vsym \geq 0.5$) Then, with probability $p = 0.2$ (resp. $p = 0.8$), we sampled $\lookback + 1$ values from the uniform distribution $U(0.5, 1)$ (resp. $U(0, 0.5)$). This assumed risk scores are computed daily and includes the present day. We generated the times of these risk scores by sampling a time offset $\tsym_{\text{off}} \sim U(0\text{s}; 86,400\text{s})$ for each user such that $\tsym_d = \tsym_{\text{now}} + \tsym_{\text{off}} - d~\text{days}$, where $d \in [0, \lookback]$. To generate contact times, we followed the same procedure for risk scores, except that we randomly sampled one of the $\lookback + 1$ times and used that as the contact time.

We evaluated various transmission rates and send coefficients: $(\scoeff, \trate) \in \{0.1, 0.2, \ldots, 1\} \times \{0.1, 0.2, \ldots, 0.9\}$. For all $(\scoeff, \trate)$ pairs, we used $n = 5,000$ and $\nactors = 2$.

To measure the scalability of risk propagation, we considered $n \in [10^2, 10^4]$ users in increments of 100 and collected 10 iterations for each $n$. The number of actors we used depended on $n$ such that $\nactors(n) = 1$ if $n < 10^3$ and $\nactors(n) = 2$ otherwise. Increasing $\nactors$ for our choice of $n$ did not improve the performance due to the communication overhead.

\subsubsection{Real-World Networks}
We analyzed the efficiency of risk propagation on the following real\hyp{}world contact networks that were collected through the SocioPatterns collaboration: a high school (Thiers13) \cite{Fournet2014}, a workplace (InVS15), and a scientific conference (SFHH) \cite{Genois2018}. Because of the limited availability of real-world, large\hyp{}scale contact networks, we only evaluated risk\hyp{}propagation scalability on synthetic networks.

To ensure that all initial risk scores were propagated, we shifted contact times forward by $\tsym_{\text{now}}$ and used ($\tsym_{\text{now}} - 1$ day) when generating risk\hyp{}score times. In this way, we ensured that the most recent risk score was still older than the first contact time. Risk-score values were generated in the same manner as described in \Cref{sec:synthetic-eval} with the exception that we only generated one score. Lastly, we repeated each experiment 10 times and report the average of the results.

\subsection{Results}
\subsubsection{Efficiency}
\begin{figure*}[tbh]
\centering
\begin{tikzpicture}
\begin{groupplot}[
	group style={
		group size=3 by 1,
		xlabels at=edge bottom,
		ylabels at=edge left
	},
	boxplot,
	table/col sep=comma,
	boxplot/draw direction=y,
	xtick distance=2,
	scaled x ticks={base 10:-1},
	width=0.35\linewidth,
	height=0.29\textwidth,
	ymin=-0.1,
	ytick distance=0.2,
	xtick scale label code/.code={},
	xlabel={Send coefficient}
]
	\nextgroupplot[table/y=NormalizedUpdates,title={Normalized update count}]
	\foreach \t in {1,...,10} {
		\addplot[color=black] table[only if={entry of SendTolerance is \t}]{anc/tolerance-updates.csv};
	}%
	\nextgroupplot[table/y=NormalizedRuntimeInSeconds,title={Normalized runtime}]
	\foreach \t in {1,...,10} {
		\addplot[color=black] table[only if={entry of SendTolerance is \t}]{anc/tolerance-runtime.csv};
	}%
	\nextgroupplot[table/y=NormalizedMessages,title={Normalized message count}]
	\foreach \t in {1,...,10} {
		\addplot[color=black] table[only if={entry of SendTolerance is \t}]{anc/tolerance-messages.csv};
	}
\end{groupplot}%
\end{tikzpicture}%
\caption{Effects of send coefficient on efficiency. All dependent variables are normalized across networks and transmission rates. Update count is the number of users whose exposure score was different from their initial score; a higher normalized value indicates better accuracy. Runtime is the duration of risk propagation between \cref{item:send} and \cref{item:collect} of \Cref{alg:rp-main}; a smaller runtime indicates improved performance. Message count is the number of messages sent by actors; a lower count indicates lower communication overhead.}
\label{fig:efficiency}
\end{figure*}
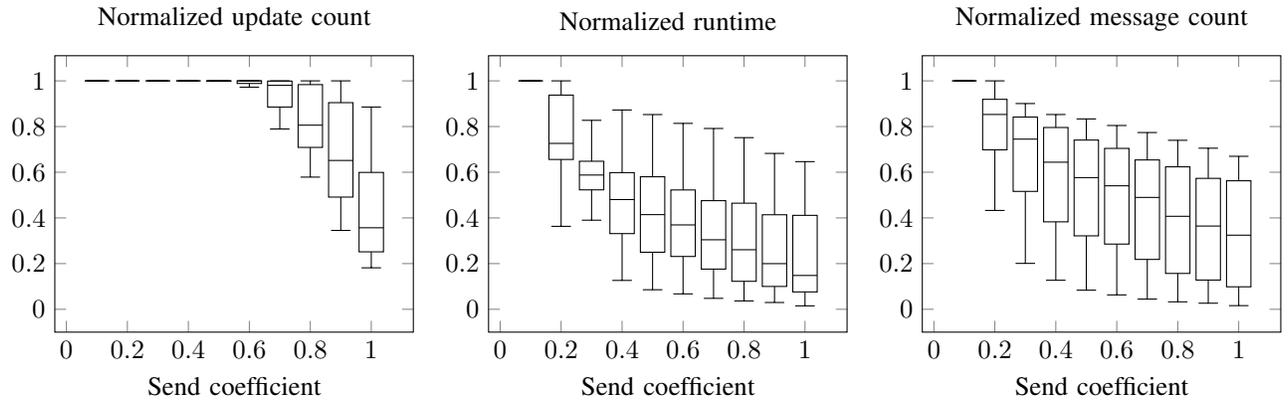

Prior to measuring scalability and real\hyp{}world performance, we observed how the send coefficient $\scoeff$ and transmission rate $\trate$ affect risk\hyp{}propagation efficiency. As ground truth for a given $\trate$, we used the maximum update count. \Cref{fig:efficiency} indicates that $\scoeff = 0.6$ permitted 99\% of the possible updates. Beyond $\scoeff = 0.6$, however, the transmission rate had considerable impact, regardless of the network. As noted in \Cref{sec:reachability}, the send coefficient quantifies the trade\hyp{}off between accuracy and efficiency; $\scoeff = 0.6$ optimized for both criteria. Herein, ``default parameters'' refers to $\trate = 0.8$ and $\scoeff = 0.6$.

Unlike the update count, \Cref{fig:efficiency} shows a more variable relationship with respect to runtime and message count. While, in general, $\trate$ (resp. $\scoeff$) had a direct (resp. inverse) relationship with runtime and message count, the network topology seems to have an impact. Namely, LFRGs displayed less variability across $\trate$ and $\scoeff$ values than RGGs and CSFGs, which is the cause for the large interquartile ranges. Therefore, it is useful to consider the lower quartile $Q_1$, the median $Q_2$, and the upper quartile $Q_3$. With default parameters, risk propagation is more efficient with $(Q_1, Q_2, Q_3) = (0.13, 0.13, 0.46)$ normalized runtime and $(Q_1, Q_2, Q_3) = (0.13, 0.15, 0.44)$ normalized message count.

\subsubsection{Message Reachability}
To validate the accuracy of \Cref{eq:estreach}, we collected values of \Cref{eq:reach} and \Cref{eq:estreach} for real\hyp{}world and synthetic contact networks. For the latter set of networks, we observed reachability while sweeping across values of $\trate$ and $\scoeff$.

To measure the accuracy of \Cref{eq:estreach}, let the \emph{message\hyp{}reachability ratio} (MRR) be defined as
\begin{equation}\label{eq:mrr}
    \mrr(u) = \frac{\reach(u)}{\estreach(u)}.
\end{equation}

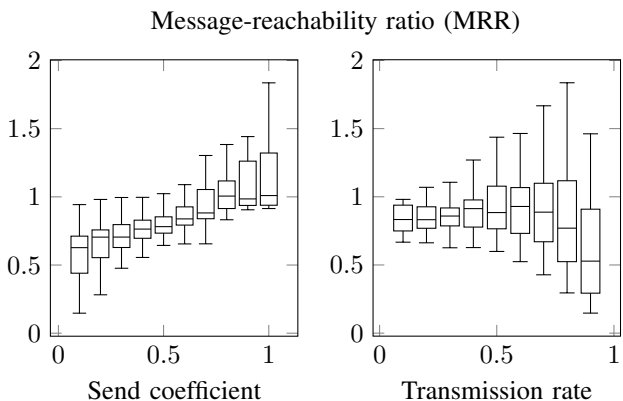
\begin{figure}[tbh]
\centering
\begin{tikzpicture}
\begin{groupplot}[
	group style={
	group size=2 by 1,
	xlabels at=edge bottom,
	ylabels at=edge left
	},
	boxplot,
	table/col sep=comma,
	boxplot/draw direction=y,
	xtick distance=5,
	scaled x ticks={base 10:-1},
	width=0.55\linewidth,
	height=0.29\textwidth,
	ytick distance=0.5,
	xtick scale label code/.code={}
]
	\nextgroupplot[table/y=RatioValue,xlabel={Send coefficient}]
	\foreach \t in {1,...,10} {
		\addplot[color=black] table[only if={entry of SendTolerance is \t}]{anc/ratio-tolerance.csv};
	}%
	\nextgroupplot[table/y=RatioValue,xlabel={Transmission rate}]
	\foreach \t in {1,...,9} {
		\addplot[color=black] table[only if={entry of Transmission is \t}]{anc/ratio-transmission.csv};
	}%
\end{groupplot}
\node (title) at ($(group c1r1.north)!0.5!(group c2r1.north)$) [above, yshift=\pgfkeysvalueof{/pgfplots/every axis title shift}] {Message\hyp{}reachability ratio (MRR)};
\end{tikzpicture}
\caption{Effects of send coefficient and transmission rate on the MRR. Independent variables are grouped across networks. A MRR above (resp. below) 1 indicates understimation (resp. overstimation).}
\label{fig:ratio}
\end{figure}

Overall, \Cref{eq:estreach} is a good estimator of \Cref{eq:reach}. Across all synthetic networks, \Cref{eq:estreach} modestly underestimated \Cref{eq:reach} with MRR quartiles $(Q_1, Q_2, Q_3) = (0.71, 0.84, 0.98)$. For $\trate = 0.8$ and $\scoeff = 0.6$, $(Q_1, Q_2, Q_3) = (0.52, 0.77, 1.12)$ and $(Q_1, Q_2, Q_3) = (0.79, 0.84, 0.93)$, respectively. \Cref{tab:reachability} provides mean values of \Cref{eq:mrr} for both synthetic and real\hyp{}world networks. \Cref{fig:ratio} indicates that a moderate $\scoeff$ produces a more stable MRR, with low (resp. high) $\scoeff$ values underestimating (resp. overestimating) \Cref{eq:reach}. For values of $\trate$, \Cref{eq:mrr} tends to decrease with increasing $\trate$, but also exhibits larger interquartile ranges.

Because \Cref{eq:estreach} does not account for the temporality constraints \Cref{eq:contact-const} and \Cref{eq:time-const}, it does not perfectly estimate \Cref{eq:reach}. With lower (resp. higher) $\scoeff$ (resp. $\trate$), \Cref{eq:estreach} suggests higher message reachability. However, because a message is only passed under certain conditions (see \Cref{alg:rp-msg}), this causes \Cref{eq:estreach} to overestimate \Cref{eq:reach}. While \Cref{eq:estreach} is a theoretical upper bound on \Cref{eq:reach}, it is possible for it to underestimate \Cref{eq:reach} if the chosen value of $\ival(v)$ overestimates the true value of $\ival(v)$. When computing \Cref{eq:mrr} for \Cref{fig:ratio}, we used the mean value of $\ival(v)$ across all users, so $\mrr(u) > 1$ in some cases.
\begin{table}[tbh]
\caption{Message\hyp{}reachability ratio for synthetic and real\hyp{}world contact networks ($\trate = 0.8$, $\scoeff = 0.6$). Synthetic (resp. real\hyp{}world) ratios are averaged across parameter combinations (resp. runs).}
\label{tab:reachability}
\centering
\begin{tabular}{lccc}
	\toprule
	& \multicolumn{3}{c}{$\mrr(u) \pm 1.96 \cdot \text{SE}$} \\
	\midrule
	\emph{Synthetic} & LFR & RGG & CSFG \\
	{\bfseries 0.85 $\pm$ 0.08} & 0.88 $\pm$ 0.14 & 0.74 $\pm$ 0.12 & 0.90 $\pm$ 0.14 \\
	\midrule
	\emph{Real-world} & Thiers13 & InVS15 & SFHH \\
	{\bfseries 0.60 $\pm$ 0.01} & 0.58 $\pm$ 0.01 & 0.63 $\pm$ 0.01 & 0.60 $\pm$ 0.01 \\
	\bottomrule
\end{tabular}
\end{table}
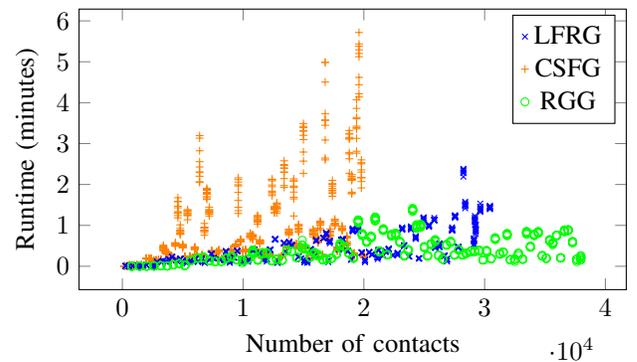
\begin{figure}[tbh]
\centering
\begin{tikzpicture}
\begin{axis}[
	width=\linewidth,
	height=0.6\linewidth,
	xlabel={Number of contacts},
	ylabel={Runtime (minutes)},
	ytick distance = 60,
	scaled y ticks={real:60},
	ytick scale label code/.code={}
]
	\addplot[
		scatter,
		only marks,
		scatter src=explicit symbolic,
		scatter/classes={
		1={mark=x,blue},
		2={mark=+,orange},
		3={mark=o,draw=green}
		},
		mark size=1.5pt
	] table [col sep=comma,x=Edges,y=RuntimeInSeconds,meta=Graph]{anc/scalability.csv};
	\legend{LFRG,CSFG,RGG}
\end{axis}%
\end{tikzpicture}%
\caption{Runtime of risk propagation on synthetic networks containing 100--10,000 users and approximately 200--38,000 contacts.}
\label{fig:runtime}
\end{figure}

\subsubsection{Scalability}
\Cref{fig:runtime} describes the runtime behavior of risk propagation. The runtime of CSFGs requires further investigation. A linear regression fit explains ($R^2 = 0.52$) the runtime of LFRGs and RGGs with a slope $m = (1.1 \pm 0.1) \cdot 10^{-3}$ s/contact and intercept $b = 4.3 \pm 1.6$s ($\pm 1.96 \cdot \text{SE}$).

\section{Conclusions}
Despite the improved formulation of risk propagation that this work provides, limitations still exist regarding its design. Unlike its proposed distributed extension \cite{Ayday2021}, this work still assumes risk propagation is an offline algorithm, which has communication complexity and privacy drawbacks. In a centralized setting, whether colocated or distributed, we must retrieve all user data to construct the contact network, but centralized data aggregation has inherent privacy issues \cite{Ayday2021}. Also different from its distributed extension is that this work associates an actor with a set of users. Because the message\hyp{}passing semantics are at the user level, subnetwork actors introduces unnecessary design complexity, especially in an online setting in which the network is dynamic.

As future work, we will develop a form of risk propagation that is online, decentralized, and asynchronous. Such a design will align with the principles of self-soverign identity \cite{Preukschat2021} and mobile\hyp{}crowdsensing applications \cite{Capponi2019} that incentivize user engagement in exchange for personal utility (e.g., cryptocurrency, knowledge of infection risk). Additionally, we will study how concurrency, network topology, and the distributions of risk-score values, risk-score times, and contact times affect risk propagation behavior. This line of evaluation aims to understand how temporal\hyp{}network dynamics relate to concurrency, topology, and message passing \cite{Masuda2021}. While the current urgency of digital contact tracing may be low, it remains important to consider effective, privacy\hyp{}preserving solutions that will help mitigate future pandemics.

\section*{Acknowledgments}
\addcontentsline{toc}{section}{Acknowledgments}
This work made use of the High Performance Computing Resource in the Core Facility for Advanced Research Computing at Case Western Reserve University.

\bibliographystyle{IEEEtran}
\bibliography{refs}

\end{document}